# Mechanical stresses estimation in silicon and glass bonded at elevated temperature

Sinev L. S.

During electrostatic bonding, also known as anodic bonding, silicon is bonded to glass by applying an external voltage and simultaneous heating to temperatures of 200...450 °C. While cooling to working temperature after bonding happened pieces are mutually deformed. Due to linear thermal expansion coefficients mismatch of anodically bonded glass and silicon samples an internal stress state is generated. Such stresses are called thermal mismatch stresses. The aim of this paper is a determination of technological and design solutions to achieve minimal thermal mismatch stresses in resulting bond.

The nonlinear dependence of linear thermal expansion coefficients of bonded samples' materials on temperature makes it difficult to minimize thermal mismatch stresses by chosing materials with close average thermal expansion coefficients in particular temperature range. To assess means of lowering thermal mismatch stress in this paper two different ways to describe assembly are used: two thin bonded layers and multilayered composite material.

Based on properties of two brands of glass (LK5, Borofloat 33) and silicon used with described mathematical models thermal mismatch stresses at temperature $T_w$ in samples bonded at several different temperatures $T_b$ are evaluated. Bonded silicon surface stress dependence of glass to silicon wafer thickness ratio is evaluated. Based on such evaluations one can say that by varying thickness of glass bonded to silicon one can obtain zero thermal mismatch stress at a particular depth of material or obtain stress of some defined value at this depth.

Models of assembly description used in this paper can be used to optimize anodic bonding process parameters. Such usage aimed to minimize thermal mismatch stresses at device working temperatures is presented in this paper.

# Оценка механических напряжений в соединённых при повышенной температуре кремнии и стекле

Синев Л. С.

УДК 621.3.049.77::621.3.082.61::536.413.2

## Введение

В свете тенденции к миниатюризации измерительных и управляющих приборов и развития соответствующих технологий всё большую важность приобретает снижение взаимного влияния материалов, контактирующих внутри одного чувствительного элемента микроэлектромеханической системы. В частности это относится к механическим напряжениям, возникающим после сборки кремниевых и стеклянных деталей микромеханических приборов электростатическим соединением.

Электростатическое соединение является распространённой сборочной операцией в микросистемной технике. В данном процессе кремний соединяется со стеклом посредством приложения внешней разности потенциалов и одновременного нагрева до температур 200…450 °C [1, 2, 3]. В русскоязычной литературе этот процесс имеет также следующие названия: анодная посадка, анодное сращивание, термоэлектростимулированное соединение, электростатическая сварка, электроадгезионное соединение, электродиффузионная сварка, электрохимическая сварка в твёрдой фазе, сварка в электростатическом поле. В англоязычной литературе устоялись следующие синонимичные названия этого процесса: anodic bonding, field assisted bonding, electrostatic bonding, Mallory bonding, electrostatic welding.

До начала нагрева стеклянная и кремниевая детали имеют одинаковые размеры, при нагреве детали расширяются неравномерно и при температуре соединения $T_b$ имеют отличающиеся размеры. После соединения детали, охлаждаясь до рабочей температуры $T_w$, взаимно деформируются. В результате соединения образуются коэффициентные напряжения, то есть напряжения, возникающие вследствие разности значений коэффициентов теплового линейного расширения (КТЛР) стекла и кремния [4].

Целью данной работы является определение технологических и конструктивных решений для обеспечения соединения с минимальными коэффициентными напряжениями.

## Анализ способов оценки

Истинным коэффициентом теплового линейного расширения называется отношение изменения линейного размера тела, делённого на его начальный размер, к малому из-менению температуры, вызвавшему изменение размера тела [5]:

$$\alpha = \frac{1}{l_0} \cdot \frac{dl}{dT},$$

где $\alpha$ — истинный коэффициент теплового расширения, 1/°C; $l_0$ — начальный линейный размер тела, м; $dl$ — изменение линейного размера тела, м; $dT$ — малое изменение температуры, вызвавшее изменение размера тела, °C.

В мировой практике известно несколько способов оценки таких напряжений разной степени сложности. Самой простой оценкой коэффициентных напряжений является расчёт по следующей формуле [6]:

$$\sigma = E(\alpha_1 - \alpha_2)\Delta T,$$

где $\sigma$ — механические напряжения в детали, вызванные разницей между коэффициентами теплового расширения материалов, Па; $E$ — модуль упругости первого рода (модуль Юнга) материала, в котором исследуются напряжения, Па; $\alpha_1$, $\alpha_2$ — средние коэффициенты теплового линейного расширения каждого из пары соединяемых материалов, 1/°C; $\Delta T$ — разница между температурой соединения материалов и температурой, при которой исследуются коэффициентные напряжения, °C.

Эта формула применима при следующих допущениях: деформации полностью упругие, разница в КТЛР компенсируется только за счёт материала детали, в которой определяют напряжения. В этой формуле никак не учитываются толщины материалов и нелинейный характер зависимости КТЛР материалов от температуры.

По следующей формуле оценивают напряжения, возникающие на свободной поверхности кремния (верхней пластины) после сборки перевёрнутого кристалла с текстолитовым основанием [7]. Используется модель двухслойного материала, находящегося под тепловой нагрузкой. Каждый из слоёв считается изотропным:

$$\sigma = \frac{\varepsilon_m M_1 hm(2 + 3h + h^3 m)}{1 + hm(4 + 6h + 4h^2)},$$

$$M_1 = \frac{E_1}{1 - \mu_1},$$

$$M_2 = \frac{E_2}{1 - \mu_2},$$

$$m = \frac{M_2}{M_1},$$

$$h = \frac{h_2}{h_1},$$

$$\varepsilon_m = (T_w - T_b)(\alpha_1 - \alpha_2),$$

где $h_1$, $h_2$ — толщины верхней и нижней пластин, соответственно, м; $E_1$, $E_2$ — модули упругости первого рода верхней пластины и нижней пластины, соответственно, Па;

$\mu_1$, $\mu_2$ — коэффициенты Пуассона верхней пластины и нижней пластины, соответственно; $T_b$ — температура соединения пластин, °C; $T_w$ — температура определения напряжений, °C; $\alpha_1$, $\alpha_2$ — средние коэффициенты теплового линейного расширения верхней пластины и нижней пластины в рассматриваемом интервале температур, соответственно, 1/°C; $\varepsilon_m$ — относительная деформация вызванная разницей коэффициентов теплового расширения соединённых пластин.

Помимо того, что эта модель предполагает оценку напряжений только на свободных поверхностях соединяемых материалов, в ней не учитывается нелинейность КТЛР.

## Предлагаемые модели оценки

Нелинейная зависимость КТЛР соединяемых деталей от температуры не позволяет минимизировать коэффициентные напряжения путём подбора материалов с близкими средними КТЛР. Для более точной оценки таких напряжений предлагается использовать модели, описываемые далее.

В [8] была представлена модель двух тонких слоёв для случая соединения кремния со стеклом:

$$\sigma_g(T) = \frac{E_g E_{si} h_{si}}{E_{si} h_{si} + E_g h_g} \int\limits_{T_b}^{T_w} (\alpha_{si}(T) - \alpha_g(T))\,\mathrm{d}T,$$

$$\sigma_{si}(T) = \frac{E_g h_g E_{si}}{E_{si} h_{si} + E_g h_g} \int\limits_{T_b}^{T_w} (\alpha_{si}(T) - \alpha_g(T))\,\mathrm{d}T, \qquad (1)$$

где $\sigma_g$, $\sigma_{si}$ — коэффициентные напряжения в деталях из стекла и кремния, соответственно, соединённых при температуре $T_b$, °C, возникающие при температуре $T_w$, °C, растягивающие в стекле и сжимающие в кремнии, Па; $E_g$, $E_{si}$ — модули упругости первого рода стекла и кремния, соответственно, Па; $\alpha_g$, $\alpha_{si}$ — истинные КТЛР стекла и кремния, соответственно, 1/°C; $h_g$, $h_{si}$ — толщины соединяемых стекла и кремния, соответственно, м.

В рамках этой модели считаем обе соединяемых детали сплошными, однородными, изотропными и непрерывными. Используем допущение, что нагрев деталей равномерен и источник тепла расположен вне области соединения. Также считаем, что область соединения представляет собой плоскость. Изменения размеров рассматриваем в плоскости перпендикулярной плоскости соединения. Считаем, что деформации и напряжения в области соединения равны деформациям и напряжениям во всей детали. Влияние краевых эффектов и разницы в коэффициентах теплопроводности материалов исключаем из рассмотрения. Исходим из того, что изгиб деталей под действием возникающих деформаций пренебрежимо мал.

Чтобы учесть распределение коэффициентных напряжений по толщине соединяемых материалов, воспользуемся теорией слоистых композитов [9, 10, 11]. Рассмотрим соединённые детали как многослойный композиционный материал. В качестве координатной плоскости $xy$ примем срединную плоскость пластины, то есть плоскость, лежащую до нагружения пластины посередине между её верхней и нижней поверхностями. Считаем, что эта плоскость не меняет своего положения в процессе нагружения. Положительным направлением оси $z$ будем считать направление вниз. Пластина состоит из произвольного числа слоёв, соединённых друг с другом. Для каждого слоя справедлив закон Гука. Предполагаем, что слои не оказывают сдавливающего воздействия один на другой. Толщину пластины считаем неизменной. Используем допущения, что нагрев пластины равномерен, что длина и ширина пластины значительно превышают её толщину. Изменение жёсткости рассматриваемых материалов считаем незначительным. Влияние краевых эффектов и разницы в коэффициентах теплопроводности материалов исключаем из рассмотрения. Также не учитываем изменение свойств стекла, связанное с переносом ионов в результате проведения процесса электростатического соединения [12]. Положительными напряжениями считаем напряжения растяжения в материале.

На рис. 1 представлена иллюстрация применяемой модели слоистого композита. В этой модели $t$ — толщина многослойной пластины; $1, 2, \ldots, k, \ldots, n$ — номер слоя.

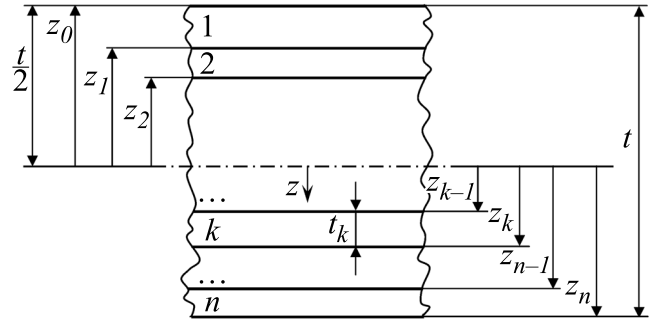

**Рис. 1.** Иллюстрация модели слоистого композита

Уравнение для напряжений в каждом слое при механическом и тепловом нагружении:
$$\boldsymbol{\sigma} = \mathbf{Q}(\boldsymbol{\varepsilon} - \boldsymbol{\varepsilon}^T), \qquad (2)$$
$$\boldsymbol{\varepsilon}^T = \int\limits_{T_b}^{T_w} \boldsymbol{\alpha}(T)\mathrm{d}T,$$
$$\boldsymbol{\varepsilon} = \boldsymbol{\varepsilon}^0 + z\boldsymbol{\kappa},$$

где $\boldsymbol{\sigma}$ — вектор напряжений, Па; $\mathbf{Q}$ — преобразованная матрица жёсткости каждого слоя, Па; $\boldsymbol{\varepsilon}$ — вектор индуцированных деформаций (растяжения), вызванных механической нагрузкой; $\boldsymbol{\varepsilon}^T$ — вектор индуцированных деформаций (растяжения), вызванных тепловой нагрузкой; $T_b$ — температура соединения, °C; $T_w$ — рабочая температура, °C; $\boldsymbol{\varepsilon}^0$ — относительное удлинение срединной поверхности многослойной композитной пластины (по осям); $\boldsymbol{\kappa}$ — радиус кривизны срединной поверхности многослойной композитной пластины (по осям), 1/м; $z$ — расстояние, измеряемое от срединной поверхности, м.

Матрицы **Q** получены из матриц жёсткости в соответствии с формулами поворота системы координат [13] так, чтобы они отражали упругие свойства слоёв в принятых нами направлениях координатных осей.

Поскольку принято допущение, что размеры пластины много больше её толщины, то далее напряжённое состояние многослойной пластины будет рассматриваться как плоское напряжённое состояние [13]. Таким образом, векторы и матрицы в уравнениях будут представлены в сокращённой форме за счёт отбрасывания незадействованных компонентов (сокращения до размера $3 \times 3$).

Учитывая вышесказанное, уравнение (2), может быть записано в матричном представлении следующим образом:

$$\begin{pmatrix} \sigma_x \\ \sigma_y \\ \sigma_{xy} \end{pmatrix} = \begin{pmatrix} Q_{11} & Q_{12} & Q_{16} \\ Q_{12} & Q_{22} & Q_{26} \\ Q_{16} & Q_{26} & Q_{66} \end{pmatrix} \left( \begin{pmatrix} \varepsilon_x^0 \\ \varepsilon_y^0 \\ \varepsilon_{xy}^0 \end{pmatrix} + z \begin{pmatrix} \kappa_x \\ \kappa_y \\ \kappa_{xy} \end{pmatrix} - \begin{pmatrix} \varepsilon_x^T \\ \varepsilon_y^T \\ \varepsilon_{xy}^T \end{pmatrix} \right),$$

где $\sigma_{ij}$ — элементы вектора напряжений, Па; $Q_{ij}$ — элементы преобразованной матрицы жёсткости каждого слоя, Па; $\varepsilon_{ij}^0$ — элементы вектора относительного удлинения срединной поверхности многослойной композитной пластины; $\kappa_{ij}$ — элементы векторного представления радиуса кривизны срединной поверхности многослойной композитной пластины, 1/м; $\varepsilon_{ij}^T$ — элементы вектора индуцированных деформаций, вызванных тепловой нагрузкой.

Результирующие силы и моменты (обобщённые силовые факторы), воздействующие на пластину, определяются посредством интегрирования уравнения (2) по всей толщине пластины:

$$\mathbf{N} = \int_t \boldsymbol{\sigma} \mathrm{d}z,$$

$$\mathbf{M} = \int_t \boldsymbol{\sigma} z \mathrm{d}z,$$

где $t$ — толщина многослойной пластины, м; **N** — результирующая нагрузка, отнесённая к единице длины линий, ограничивающих элемент рассматриваемой поверхности [14], Н/м; **M** — результирующий момент, отнесённый к единице длины линий, ограничивающих элемент рассматриваемой поверхности, Н.

В рассматриваемом случае, когда жёсткость каждого слоя **Q** неизменна по толщине слоя, можно записать:

$$A_{ij} = \sum_{k=1}^{n} (Q_{ij})_k (z_k - z_{k-1}),$$

$$B_{ij} = \frac{1}{2} \sum_{k=1}^{n} (Q_{ij})_k (z_k^2 - z_{k-1}^2),$$

$$D_{ij} = \frac{1}{3} \sum_{k=1}^{n} (Q_{ij})_k (z_k^3 - z_{k-1}^3),$$

где $A_{ij}$ — элементы матрицы жёсткости при растяжении (мембранной жёсткости) [9, 14], Н/м; $B_{ij}$ — элементы матрицы жёсткости изгиб-растяжение (смешанной жёсткости), Н; $D_{ij}$ — элементы матрицы жёсткости при изгибе (изгибной жёсткости), Н·м; $z_k$ — расстояние до текущего слоя, измеряемое от срединной поверхности (см. рис. 1).

Тогда взаимосвязь нагрузок и деформаций может быть представлена в уравнениях в матричной форме [9, 10]:

$$\left(\frac{\mathbf{N}}{\mathbf{M}}\right) = \left(\begin{array}{c|c} \mathbf{A} & \mathbf{B} \\ \hline \mathbf{B} & \mathbf{D} \end{array}\right) \left(\frac{\boldsymbol{\varepsilon}^0}{\boldsymbol{\kappa}}\right) - \left(\frac{\mathbf{N}^T}{\mathbf{M}^T}\right), \qquad (3)$$

где $\mathbf{N}^T$ — усилие, вызванное тепловым воздействием, отнесённое к единице длины линий, ограничивающих элемент рассматриваемой поверхности, Н/м; $\mathbf{M}^T$ — момент силы, вызванный тепловым воздействием, отнесённый к единице длины линий, ограничивающих элемент рассматриваемой поверхности, Н.

Преобразовав запись, получим:

$$\begin{pmatrix} N_x \\ N_y \\ N_{xy} \end{pmatrix} = \begin{pmatrix} A_{11} & A_{12} & A_{16} \\ A_{12} & A_{22} & A_{26} \\ A_{16} & A_{26} & A_{66} \end{pmatrix} \begin{pmatrix} \varepsilon_x^0 \\ \varepsilon_y^0 \\ \varepsilon_{xy}^0 \end{pmatrix} + \begin{pmatrix} B_{11} & B_{12} & B_{16} \\ B_{12} & B_{22} & B_{26} \\ B_{16} & B_{26} & B_{66} \end{pmatrix} \begin{pmatrix} \kappa_x \\ \kappa_y \\ \kappa_{xy} \end{pmatrix} - \begin{pmatrix} N_x^T \\ N_y^T \\ N_{xy}^T \end{pmatrix}, \qquad (4)$$

$$\begin{pmatrix} M_x \\ M_y \\ M_{xy} \end{pmatrix} = \begin{pmatrix} B_{11} & B_{12} & B_{16} \\ B_{12} & B_{22} & B_{26} \\ B_{16} & B_{26} & B_{66} \end{pmatrix} \begin{pmatrix} \varepsilon_x^0 \\ \varepsilon_y^0 \\ \varepsilon_{xy}^0 \end{pmatrix} + \begin{pmatrix} D_{11} & D_{12} & D_{16} \\ D_{12} & D_{22} & D_{26} \\ D_{16} & D_{26} & D_{66} \end{pmatrix} \begin{pmatrix} \kappa_x \\ \kappa_y \\ \kappa_{xy} \end{pmatrix} - \begin{pmatrix} M_x^T \\ M_y^T \\ M_{xy}^T \end{pmatrix}, \qquad (5)$$

где $N_{ij}$ — элементы вектора результирующей нагрузки, отнесённой к единице длины линий, ограничивающих элемент рассматриваемой поверхности, Н/м; $M_{ij}$ — элементы вектора результирующего момента, отнесённого к единице длины линий, ограничивающих элемент рассматриваемой поверхности, Н; $N_{ij}^T$ — элементы векторной записи усилия, вызванного тепловым воздействием, отнесённого к единице длины линий, ограничивающих элемент рассматриваемой поверхности, Н/м; $M_{ij}^T$ — элементы векторной записи момента силы, вызванного тепловым воздействием, отнесённого к единице длины линий, ограничивающих элемент рассматриваемой поверхности, Н.

Силы и моменты, вызванные тепловым воздействием, определяются следующими уравнениями.

$$\mathbf{N}^T = \int_t \mathbf{Q} \boldsymbol{\varepsilon}^T \mathrm{d}z, \qquad (6)$$

$$\mathbf{M}^T = \int_t \mathbf{Q} \boldsymbol{\varepsilon}^T z \mathrm{d}z. \qquad (7)$$

В рассматриваемом нами случае внешнее механическое нагружение отсутствует, поэтому уравнение (3) можно переписать следующим образом:

$$\left(\frac{\mathbf{N}^T}{\mathbf{M}^T}\right) = \left(\begin{array}{c|c} \mathbf{A} & \mathbf{B} \\ \hline \mathbf{B} & \mathbf{D} \end{array}\right) \left(\frac{\boldsymbol{\varepsilon}^0}{\boldsymbol{\kappa}}\right).$$

Уравнения (4, 5) при этом можно переписать:

$$\begin{pmatrix} N_x^T \\ N_y^T \\ N_{xy}^T \end{pmatrix} = \begin{pmatrix} A_{11} & A_{12} & A_{16} \\ A_{12} & A_{22} & A_{26} \\ A_{16} & A_{26} & A_{66} \end{pmatrix} \begin{pmatrix} \varepsilon_x^0 \\ \varepsilon_y^0 \\ \varepsilon_{xy}^0 \end{pmatrix} + \begin{pmatrix} B_{11} & B_{12} & B_{16} \\ B_{12} & B_{22} & B_{26} \\ B_{16} & B_{26} & B_{66} \end{pmatrix} \begin{pmatrix} \kappa_x \\ \kappa_y \\ \kappa_{xy} \end{pmatrix},$$

$$\begin{pmatrix} M_x^T \\ M_y^T \\ M_{xy}^T \end{pmatrix} = \begin{pmatrix} B_{11} & B_{12} & B_{16} \\ B_{12} & B_{22} & B_{26} \\ B_{16} & B_{26} & B_{66} \end{pmatrix} \begin{pmatrix} \varepsilon_x^0 \\ \varepsilon_y^0 \\ \varepsilon_{xy}^0 \end{pmatrix} + \begin{pmatrix} D_{11} & D_{12} & D_{16} \\ D_{12} & D_{22} & D_{26} \\ D_{16} & D_{26} & D_{66} \end{pmatrix} \begin{pmatrix} \kappa_x \\ \kappa_y \\ \kappa_{xy} \end{pmatrix}.$$

Преобразуем эту систему уравнений:

$$\boldsymbol{\varepsilon}^0 = \left(\mathbf{A}^{-1} + (\mathbf{A}^{-1}\mathbf{B})(\mathbf{D} - \mathbf{B}\mathbf{A}^{-1}\mathbf{B})^{-1}(\mathbf{B}\mathbf{A}^{-1})\right)\mathbf{N}^T - (\mathbf{A}^{-1}\mathbf{B})(\mathbf{D} - \mathbf{B}\mathbf{A}^{-1}\mathbf{B})^{-1}\mathbf{M}^T, \quad (8)$$

$$\boldsymbol{\kappa} = -(\mathbf{D} - \mathbf{B}\mathbf{A}^{-1}\mathbf{B})^{-1}(\mathbf{B}\mathbf{A}^{-1})\mathbf{N}^T + (\mathbf{D} - \mathbf{B}\mathbf{A}^{-1}\mathbf{B})^{-1}\mathbf{M}^T, \quad (9)$$

где $\mathbf{A}$ — матрица жёсткости при растяжении (мембранная жёсткость), Н/м; $\mathbf{B}$ — матрица жёсткости изгиб-растяжение (смешанная жёсткость), Н; $\mathbf{D}$ — матрица жёсткости при изгибе (изгибная жёсткость), Н·м.

Зависимости для определения коэффициентных напряжений в любой плоскости внутри рассматриваемой сборки параллельной срединной поверхности получаются подстановкой уравнений (6 – 9) в уравнение (2). Причём, зная расположение главных осей кремния, преобразуя значения элементов матриц жёсткости в элементы матрицы Q в соответствии с формулами поворота системы координат [13], можно определять напряжения в любых направлениях.

### Примеры применения моделей

Расчёты, результаты которых рассматриваются в этой работе, основаны на следующих данных о свойствах кремния и стекла. Экспериментально полученная зависимость КТЛР кремния приведена в [15]. Упругие свойства кремния ориентации (100) взяты из [16]. В качестве стёкол для расчётов были выбраны материалы марки Borofloat 33 и ЛК5, КТЛР которых были измерены экспериментально, а упругие свойства взяты из [17, 18]. Толщины деталей, если иное не оговорено, кремниевой — 0,42 мм, стеклянной — 0,6 мм. Графики изменения КТЛР представлены на рис. 2.

Воспользовавшись формулой (1), можно оценить, как будут меняться коэффициентные напряжения в детали из кремния в рабочем диапазоне температур прибора. Иллюстрация такого применения приведена на рис. 3 и рис. 4.

В данной модели для каждой пары соединяемых материалов величина разброса напряжений в определённом диапазоне рабочих температур остаётся неизменной при изменении температуры соединения, меняются только крайние значения напряжения

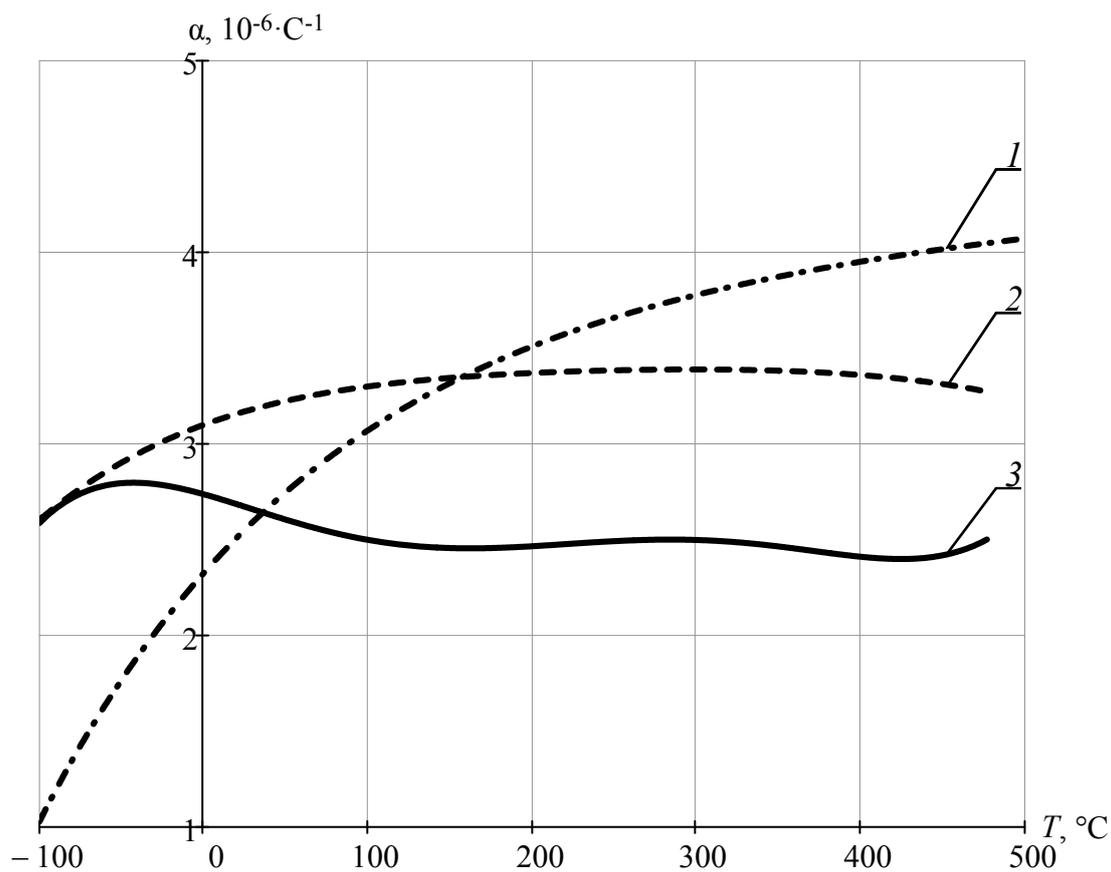

**Рис. 2.** Графики изменения КТЛР: *1* — кремний [15], *2* — ЛК5, *3* — Borofloat 33

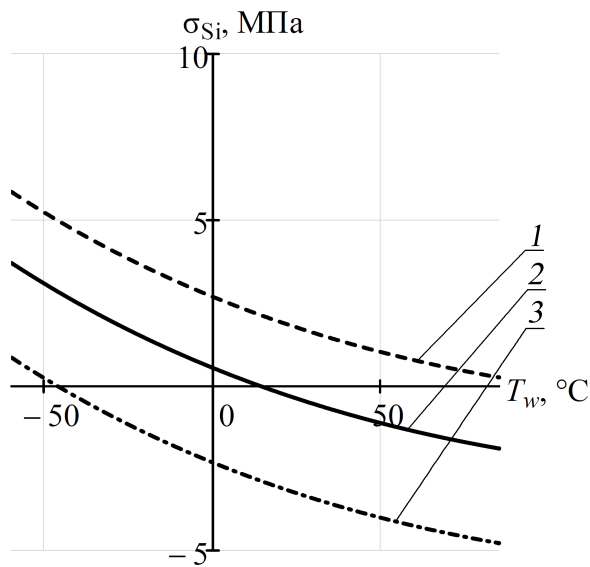

**Рис. 3.** Коэффициентные напряжения в кремнии в рабочем диапазоне температур (от минус 60 до плюс 85 °C) прибора при разных фиксированных температурах проведения процесса соединения со стеклом марки ЛК5: *1* — 250 °C, *2* — 350 °C, *3* — 450 °C

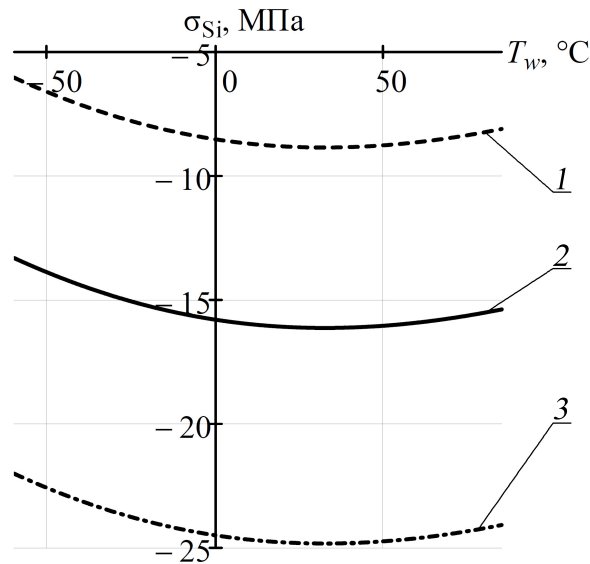

**Рис. 4.** Коэффициентные напряжения в кремнии в рабочем диапазоне температур (от минус 60 до плюс 85 °C) прибора при разных фиксированных температурах проведения процесса соединения со стеклом марки Borofloat 33: *1* — 250 °C, *2* — 350 °C, *3* — 450 °C на границе диапазона. Для представленных случаев расчётные разбросы составляют: кремний, соединённый с ЛК5 — 5,6 МПа; кремний, соединённый с Borofloat 33 — 2,8 МПа.

Поскольку не представляется возможным избежать наличия разброса коэффициентных напряжений, то имеет смысл сделать такой разброс симметричным в рабочем диапазоне, чтобы было удобно компенсировать иным способом. Для получения симметричного разброса напряжений в рабочем диапазоне следует выбрать такую температуру проведения процесса, при которой несимметричность равна нулю.

Иллюстрации применения модели многослойного композиционного материала, описанной выше уравнениями (2, 6 – 9), приведены на рис. 5 и рис. 6.

Оценка распределения коэффициентных напряжений по глубине соединяемых материалов представлена на рис. 5. Расчётные тепловые параметры, применённые для этой оценки, — рабочая температура $T_w = 20$ °C, температура соединения со стеклом ЛК5 — $T_b = 380$ °C, температура соединения со стеклом Borofloat 33 — $T_b = 300$ °C. За плоскость отсчёта координаты по оси $z$ взята плоскость соединения кремния со стеклом.

Из данной иллюстрации можно сделать вывод, что, варьируя соотношение толщины стекла и кремния, можно получить на некоторой глубине кремния коэффициентные напряжения предсказуемой величины или нулевые.

Графики на рис. 6 представляют оценку значения коэффициентных напряжений на свободной поверхности кремния при рабочей температуре в зависимости от отношения, $H$, толщины стекла к толщине кремния рассчитанные по модели многослойного композита. Расчётные тепловые параметры, применённые для этой оценки, — температура соединения со стеклом ЛК5 — $T_b = 380$ °C, температура соединения со стеклом Borofloat 33 — $T_b = 300$ °C. При толщине стекла чуть более чем в три раза превышаю-

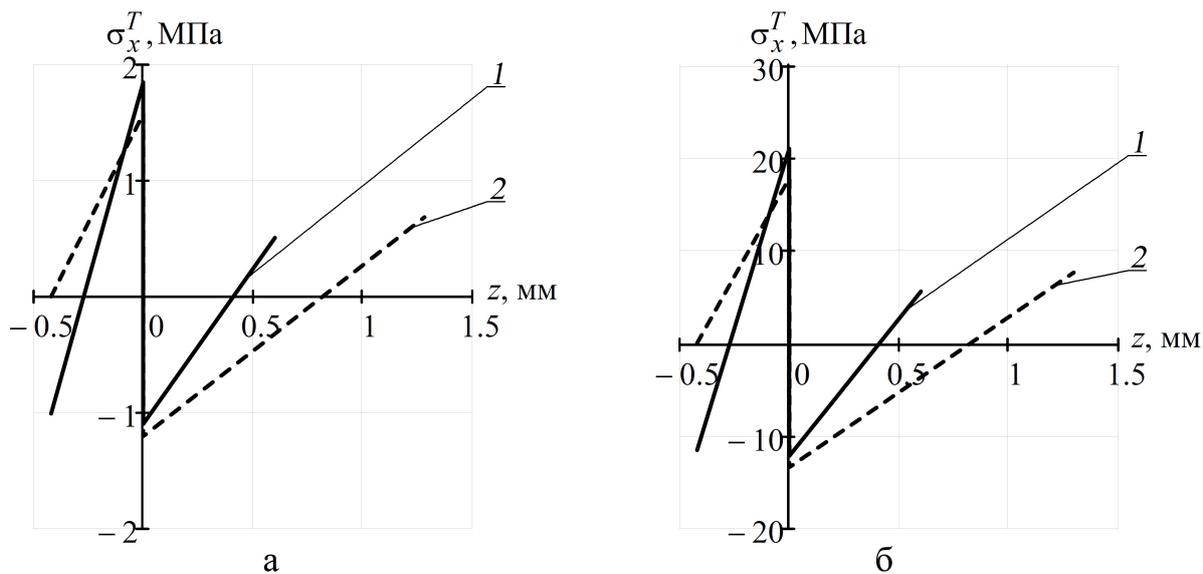

**Рис. 5.** Эпюры коэффициентных напряжений в сборке кремния со стёклами двух марок для случаев разных толщин стекла. Марка стекла: а — ЛК5, б — Borofloat 33. Толщина стекла: *1* — 0,6 мм; *2* — 1,3 мм

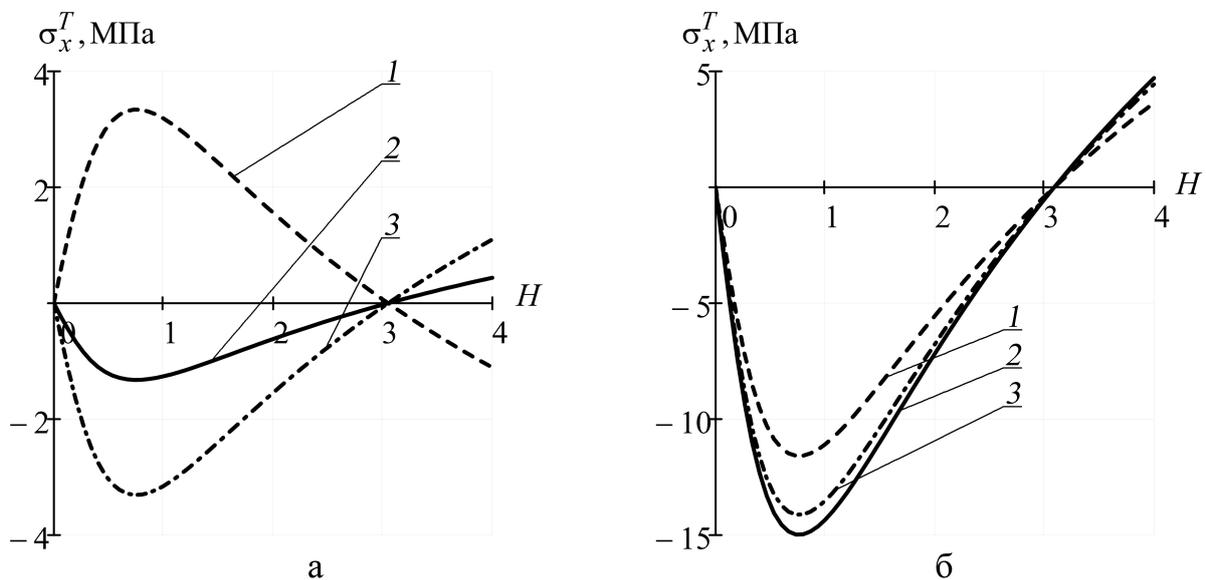

**Рис. 6.** Коэффициентные напряжения на свободной поверхности кремния, в зависимости от соотношения толщин стекла и кремния. Марка стекла: а — ЛК5, б — Borofloat 33. Рабочая температура $T_w$: *1* — минус 60 °C; *2* — 20 °C; *3* — 85 °C щей толщину кремния, согласно расчётам, на поверхности кремния будут отсутствовать коэффициентные напряжения во всём рабочем диапазоне температур.

## Заключение

В рамках данной работы рассмотрены способы оценки напряжений в сборках пластин стекла и кремния, соединённых методом электростатического соединения, проводимые в соответствии с теорией слоистых композитов. Проведены расчёты для некоторых случаев соотношений толщин материалов и температур проведения процесса. Показан способ оптимизации режима проведения процесса соединения с целью минимизации влияния коэффициентных напряжений в диапазоне рабочих температур получаемого прибора. Также сделан вывод о существовании такого соотношения толщин пластин стекла и кремния, при котором на поверхности кремния будут отсутствовать коэффициентные напряжения, независимо от температуры, при которой было проведено соединение.

## Список литературы